\documentstyle{caosp}
\begin{document}
\pubyear{1998}
\volume{27}
\firstpage{329}

\hauthor{A. Aret \& A. Sapar}

\title{Particle diffusion in atmospheres of CP stars}

\author{A. Aret \and A. Sapar}
\institute{Tartu Observatory\\ EE2444 T\~oravere, Estonia}

\date{December 30, 1997}
\maketitle

\begin{abstract}
We give concisely the formulae governing diffusion of chemical elements
and their isotopes in quiescent stellar atmospheres, due to electrostatic,
gravitational and radiation fields and to impacts between particles.
Isotope segregation of heavy elements due to
light-induced drift is emphasized.

\end{abstract}

\section{Introduction}
\label{intr}
The diffusive separation of chemical elements and their isotopes in stellar 
atmospheres  can occur only in the case of lacking
macroscopic motions, i.e. if the stellar wind and the meridional circulation
are extremely weak and there is no convective turbulence. These conditions
hold only for CP stars, and  overabundances of
heavy elements in their atmospheres can reach  several orders of magnitude.

\section{ Diffusion equations of plasma components }
The element diffusion is described with sufficient accuracy
if we take into account the collisions
between different atomic particles and the presence of external 
 gravitation, radiation and electrostatic fields
 (Sapar \& Aret, 1995). 
 Momentum transfer from particles of types $i$ to 
particles $j$ is described by
\begin{equation}\label{r7}
     {\partial ( \rho_j \vec V_j ) \over \partial t } +
      \vec\nabla P_j = 
      \rho_j \vec a_j +
      \sum_i \nu_{ij} \rho_i \vec V_i
      -\nu_j \rho_j \vec V_j  ~.      
\end{equation}
Here the collision frequency of
particle  $j$  with particles $i$ and total collision frequency
for particles $j$ are to be found from expressions
\begin{equation}\label{r1}
\nu_{ji} = n_i \int\kern-.5em\int \sigma_{ji}(v_{ji}) v_{ji} f_j f_i
 d\vec v_i d\vec v_j ~, ~~
      \nu_j = \sum_i \nu_{ji} ~,
\end{equation}
where $n_i$ is the number density, $v_{ji}$ the relative particle velocity 
and $f_i$ the undisturbed velocity distribution.
Indices indicate species of particles --- all ions are treated 
separately and electrons are included as one of them.

In the time--independent approximation we obtain from Eq.(\ref{r7})
 a system of \eject\noindent 
algebraic linear equations for diffusion velocities $V_j$:
\begin{equation}\label{r8}
      \nu_j \rho_j \vec V_j =
      \vec F_j +
      \sum_i \nu_{ij} \rho_i \vec V_i ~,~~~~~~~~\hbox{where}~~~
      \vec F_j =
      - \vec\nabla P_j +  \rho_j \vec a_j ~.
\end{equation}
 Using a buffer-gas approximation we
neglect interactions between heavy elements as small
admixtures and take into account for each of them separately  only 
interaction with buffer gas particles. 
Calculations showed, that adequate buffer gas mixture for modelling 
consists of H, He, C, N and O. 
The diffusion equations for a particular element can be derived from the
continuity eqation
\begin{equation}\label{r10}
     {\partial\rho_\varepsilon\over\partial t} +
      \vec \nabla ( \rho_\varepsilon \vec V_\varepsilon ) = 0~,~~~~~~~~
\hbox{where}~~~
     \rho_\varepsilon = \sum_{j\in\varepsilon} \rho_j~,~~~~  
     \vec V_\varepsilon = {1\over\rho_\varepsilon}
      \sum_{j\in\varepsilon} \rho_j \vec V_j ~
\end{equation}
where summing is carried out over all particle species belonging to
element $\varepsilon$.
The electron density is to be found from the condition of
gas electroneutrality.

The external acceleration of plasma particles due to  
gravity, electrostatic field, radiative
acceleration both in spectral continua and lines and due to light--induced
drift (Atutov \& Shalagin, 1988; Nasyrov \& Shalagin, 1993) has 
 the form
\begin{equation}\label{r12}
       a_j =
       \vec g_j+  { Z_j e \over m_j } \vec E + \vec a_{j,rad}~~.
\end{equation}

Electrons, since they are much lighter than
ions, tend to drift up in the atmo\-sphere,
generating an electrostatic field which blocks their escape.
The electrostatic field has almost no influence on diffusion of 
heavy elements, but for light ions its lifting force can even prevail upon
the opposite--directed gravity. 

The electrostatic field can be found from the main system of linear equations
for diffusion velocities and the condition of no electric current 
$ \sum_i n_i Z_i \vec V_i = 0 ~.$

In numerous studies (see Gonzalez et al., 1995; Michaud \& Proffitt, 1993; 
Alecian et al., 1993; Ryabchikova, 1992 and references therein) 
the radiative expelling force for the
majority of elements has been computed both for stellar atmospheres and
for their envelopes.
For line--rich
metals the expelling force highly exceeds the gravity, thus being the
dominant factor for the formation of observed metal overabundances in 
the CP stellar atmospheres..

 Some essential discrepancies between the 
 observations and theory have remained (Wahlgren et al., 1995). 
They can be partly removed when the light--induced  drift  
is also taken into account (see also LeBlanc \& Michaud, 1993).

In free--free and bound--free electron transitions the momentum 
of the photon is transferred 
to the generated ions. In  bound--bound transitions it is transferred to
particles in the upper state $u$ and the radiation flux acts on them with
a force:
\begin{equation}\label{r19}
\vec f_{ul}^r = {\pi\over c} \int n_l\,
            \sigma_{ul}(\nu)\, \vec F_\nu\, d\nu ~,~~~~~~
 \sigma_{ul}(\nu) = \sigma_{ul}^0 W(u_\nu, a)~,           
\end{equation}
where $n_l$ is the number density of particles in the lower state, 
$\vec F_\nu$ is the monochromatic flux, $\sigma_{ul}(\nu)$
is the photon absorption cross-section in the transition $l \rightarrow u$
which can be written using the Voigt function $W(u, a)$. The parameters of
the Voigt function are 
$u=(\nu -\nu_0 )/ \Delta\nu_D$ and $a=\Gamma_{ul} / (4\pi \Delta\nu_D)$.

To find the radiative force acting on a particular ion, we need to summarize 
over all transitions.
The effect of light--induced drift can be taken into account by 
substituting the Voigt function $W(u,a)$ in the radiative-force
 expression for spectral lines by
\begin{equation}\label{r20}
w(u,a)=W(u,a)-2 q D{\partial W(u,a) \over\partial u}~,~~~~
{\partial W(u,a) \over \partial u}=-{\partial W(-u,a) \over \partial u}~.
\end{equation}
The expression $q={c\sqrt{2 MkT} /h\nu}$ is the ratio of the mean momentum
of the ion species studied to the momentum of the absorbed light quantum,
$h\nu /c$, being about $10^4.$
The light-induced drift is generated due to the lower mobility of atomic
particles in the excited (upper) states resulting from larger impact
cross-sections than in the ground (lower) state, yielding thus
 lower diffusion rates (particle mobilities)
for excited states. 

The uncompensated drift rate due to the difference of
cross-sections $\sigma$ is given by the expression 
$D_1=1-\sigma_l/\sigma_u=\Delta\sigma/
\sigma_u.$ The spontaneous transitions which take place before collisions 
reduce the drift rate by factor
$D_2=1-A/(A+\nu_u)=\nu_u/(A+\nu_u)$, where $A$ is  the probability 
(frequency) of spontaneous transitions. Thus, the total reduction factor is
$D=D_1 D_2=\Delta\nu/( A+\nu_u)$.
The effective value of $2Dq$ reduces to about $10^2$ in the upper atmosphere
 and reaches to  about $10^4$ in the deeper layers.
 
 The most effective case for light-induced drift is if in 
blends, the fluxes in the blue and red wing of a spectral line
 differ essentially, especially for isotope splitting
of resonance spectral lines of heavy elements where the isotopic wavelength 
difference of spectral lines is of the order of the thermal Doppler width.
In this case  only the heaviest isotopes dominate in the atmospheres as
observed, say, by HST for Hg, Tl and Pt in the spectrum of $\chi$ Lup
(Wahlgren et al., 1995).

\end{document}